\documentstyle[aps,prd,epsf]{revtex}
\begin{document}
\draft
\title{Reconstructing the inflationary power spectrum from CMBR data}
\author{Steen Hannestad}
\address{NORDITA, Blegdamsvej 17, DK-2100 Copenhagen, Denmark}
\date{\today}

\maketitle

\begin{abstract}
The Cosmic Microwave Background Radiation (CMBR) holds information
about almost all the fundamental cosmological parameters,
and by performing a likelihood analysis of high precision CMBR fluctuation
data, these parameters can be inferred. However, this analysis relies 
on assumptions about the initial power spectrum, which is usually
taken to be a featureless power-law, $P(k) \propto k^{n_s-1}$.
Many inflationary models predict power spectra with non-power law
features. We discuss the possibility for detecting such features
by describing the power spectrum as bins in $k$-space.
This method for power spectrum reconstruction is demonstrated in
practise by performing likelihood optimization on synthetic
spectra, and the difficulties arising from reconstructing smooth
features using discontinuous bins are discussed in detail.
\end{abstract}

\pacs{PACS numbers: 98.70.Vc, 98.80.C, 98.80.-k}


\section{introduction}

Fluctuations in the Cosmic Microwave Background Radiation
(CMBR) were detected for the first time by the COBE satellite in 1992
\cite{COBE}.
Subsequently it was realised that precision measurements of the
fluctuation spectrum can be used to infer almost all of the fundamental
cosmological parameters 
\cite{boef,jungman1,jungman,tegmark1,bond1,eisenstein}. 
The method most commonly used for determining parameters from
the data is to maximise the likelihood function, $\cal L$,
over the space of model parameters to be determined, 
$\theta =  \{\Omega,\Omega_m,\Omega_b,H_0,\tau,\ldots\}$.
A large number of papers have dealt with
with this issue in great detail. However, when calculating the
theoretical CMBR power spectrum needed for the likelihood
function, a necessary input is the 
initial power spectrum of fluctuations, 
usually assumed to have been produced during
the inflationary epoch.

There exists a plethora of different inflationary models, each
having a specific prediction for the produced fluctuation spectrum.
In the simplest slow-roll models, the spectrum is described by a
simple featureless power-law $P(k) \propto k^{n_s-1}$, where
$n_s \simeq 1$ \cite{kolb}. 
However, the power spectrum could easily behave
differently, depending both on the specific form of the inflaton
potential, $V(\phi)$, and on the presence of different physical phenomena
during inflation.
A plethora of different models predicting such behaviour exists \cite{infl1}.
Examples of such phenomena are the resonant production of particles
during inflation, proposed by Chung et al.\ \cite{chung}, and the multiple inflation
model discussed by Adams, Ross and Sarkar \cite{ars97}.

In almost all existing likelihood analyses the initial power
spectrum is assumed to have the power-law form $P(k) \propto k^{n_s-1}$,
where $n_s$ is a constant 
\cite{boef,jungman1,jungman,tegmark1,bond1,eisenstein}.
This type of analysis restricts the parameter estimation in
such a way that no non-power law features can be detected.
Since there are so many different inflationary models, each with
unique predictions, it is highly desirable to have a more
model-independent way of estimating the initial power spectrum.

Such a possibility has been discussed by Souradeep et al.\ \cite{soura} and 
by Wang, Spergel and Strauss \cite{wss,wss2}. 
In both treatments, the power spectrum
was described as a featureless spectrum, but binned in $k$-space.
Using a Fisher matrix analysis, it was shown by Wang, Spergel and 
Strauss that it is possible to determine the power in each
bin to reasonable precision, so that some features could be 
detectable. In the next section we discuss the prospects for detecting
features in more detail, also using the Fisher matrix technique. 
Particularly, it is shown that there
is a trade-off between the precision to which the power in each
bin can be measured and the width of each bin.

In Section III we go on to discuss in detail how such a likelihood
analysis with a binned power spectrum is performed in practise. This analysis
highlights some of the difficulties which can be expected.
In general it is quite difficult to reconstruct a power spectrum
with smooth features using a set of discontinuous bin amplitudes.
If the binning is too coarse, the individual bins being comparable
in size or broader than the power spectrum features, the
spectrum reconstruction can easily give misleading results.
For a reasonable reconstruction it is necessary to choose a binning
which is significantly finer than the features to be detected.
However, maximising the likelihood function is a highly non-linear
optimization problem. If the number of free parameters in the fit,
$N = N_{\theta}+N_{\rm bins}$, is too large the computation time 
to retrieve the maximum likelihood becomes very long.
Nevertheless, it is demonstrated that it is computationally 
feasible to use at least 20 bins in $k$-space.

Finally, Section IV contains a discussion of the results, with
emphasis on how the techniques can be used on results from high
precision satellite experiments, like MAP and PLANCK \cite{MAP+PLANCK}.


\section{Fisher Matrix Analysis}

CMBR temperature fluctuations are usually expressed in terms of spherical harmonics
as
\begin{equation}
\frac{\Delta T}{T} (\theta,\phi) = \sum_{lm} a_{lm} Y_{lm} (\theta,\phi).
\end{equation}
From these $a_{lm}$ coefficients it is possible to construct the
power spectrum as 
\begin{equation}
C_l = \langle|a_{lm}|^2 \rangle,
\label{eq:ensemble}
\end{equation}
where the average in principle is an ensemble average. Such an ensemble
average can obviously not be performed since we have access to only
one realisation of the underlying distribution. However, in 
a universe which is isotropic the ensemble average can be 
replaced by an average over $m$-values for a given $l$ \cite{coles}.

It is possible to estimate the precision with which the cosmological
model parameters can be extracted from a given hypothetical data set.
The starting point for any parameter extraction is the vector of
data points, $x$. This can be in the form of the raw data, or in
compressed form, either as the $a_{lm}$ coefficients
or the power spectrum, $C_l$.
Each data point has contributions from both signal and noise,
$x = x_{\rm CMBR} + x_{\rm noise}$. If both signal and noise are
Gaussian distributed it is possible to build a likelihood function
from the measured data which has the following form \cite{oh}
\begin{equation}
{\cal L}(\Theta) \propto \exp \left( -\frac{1}{2} x^\dagger 
[C(\Theta)^{-1}] x \right),
\end{equation}
where $\Theta = (\Omega, \Omega_b, H_0, n, \tau, \ldots)$ is a vector
describing the given point in model parameter space and 
$C(\Theta) = \langle x x^T \rangle$ 
is the
data covariance matrix.
In the following we shall always work with data in the form of a
set of power spectrum coefficients, $C_l$.

If the data points are uncorrelated so that the data covariance matrix
is diagonal, the likelihood function can be reduced to
${\cal L} \propto e^{-\chi^2/2}$, where
\begin{equation}
\chi^2 = \sum_{l=2}^{N_{\rm max}} \frac{(C_{l, {\rm obs}}-C_{l,{\rm theory}})^2}
{\sigma(C_l)^2},
\label{eq:chi2}
\end{equation} 
is a $\chi^2$-statistics and $N_{\rm max}$ 
is the number of power spectrum data
points \cite{oh}.

The maximum likelihood is an unbiased estimator, which means that
\begin{equation}
\langle \Theta \rangle = \Theta_0.
\end{equation}
Here $\Theta_0$ indicates the true parameter vector of the underlying
cosmological model and $\langle \Theta \rangle$ is the average estimate
of parameters from maximising the likelihood function.

The likelihood function should thus peak at $\Theta \simeq \Theta_0$, and
we can expand it to second order around this value.
The first order derivatives are 
zero, and the expression is thus
\begin{equation}
\chi^2  = \chi^2_{\rm min} + \sum_{i,j}(\theta_i-\theta) \left( \sum_{l=2}^{N_{\rm max}}
\frac{1}{\sigma (C_l)^2} \left[\frac{\partial C_l}{\partial \theta_i}
\frac{\partial C_l}{\partial \theta_j} - (C_{l, {\rm obs}}-C_l)
\frac{\partial^2 C_l}{\partial \theta_i \partial \theta_j} \right]\right) (\theta_j-\theta),
\end{equation}
where $i,j$ indicate elements in the parameter vector $\Theta$.
The second term in the second derivative can be expected to be very small
because $(C_{l, {\rm obs}}-C_l)$ is in essence just a random measurement error 
which should average out. The remaining term
is usually referred to as the Fisher information matrix
\begin{equation}
F_{ij} = \frac{\partial^2 \chi^2}{\partial \theta_i \partial \theta_j} = 
\sum_{l=2}^{N_{\rm max}}\frac{1}{\sigma (C_l)^2}\frac{\partial C_l}{\partial \theta_i}
\frac{\partial C_l}{\partial \theta_j}.
\label{eq:fisher1}
\end{equation}
The Fisher matrix is closely related to the precision with which the
parameters, $\theta_i$, can be determined.
If all free parameters are to be determined from the data alone without
any priors then it follows from the Cramer-Rao inequality
\cite{kendall} that
\begin{equation}
\sigma(\theta_i) = (F^{-1})_{ii}
\end{equation}
for an optimal unbiased estimator, such as the maximum likelihood
\cite{tth}.

In general, the standard error on the $C_l$ coefficients can be written as
\begin{equation}
\sigma(C_l) = \left[\frac{2}{(2l+1)f_{\rm sky}}\right]^{1/2}(C_l + \Delta_{\rm exp}).
\end{equation}
Here, $f_{\rm sky}$ is the sky coverage and $\Delta_{\rm exp}$ is a (Gaussian) experimental
error. Notice that even for $\Delta_{\rm exp}=0$, $\sigma(C_l) \neq 0$. This derives
from the fact that the ensemble average in Eq.~(\ref{eq:ensemble}) has been replaced
by an average over $m$-values, and is usually referred to as ``Cosmic Variance'' \cite{coles}.

The CMBR is also predicted to be polarized, and this polarization power spectrum
can in principle be measured.
For scalar perturbations there are only three independent quantities: The
temperature, $C_{l,T}$, the $E$-field polarization, $C_{l,E}$, and the
temperature-polarization cross-correlation, $C_{l,C}$. Then the Fisher matrix
of Eq.~(\ref{eq:fisher1}) is instead \cite{tth}
\begin{equation}
F_{ij} = 
\sum_{l=2}^{N_{\rm max}}\sum_{X,Y}\frac{\partial C_{l,X}}{\partial \theta_i}
{\rm Cov}^{-1}(C_{l,X},C_{l,Y})\frac{\partial C_{l,X}}{\partial \theta_j}.
\end{equation}
In this case the covariance matrix,
${\rm Cov} (C_{l,X},C_{l,Y})$, is a symmetric $3\times 3$ matrix with
the following elements \cite{tth}
\begin{eqnarray}
{\rm Cov}(C_{l,T},C_{l,T}) & = & \left[\frac{2}{(2l+1)f_{\rm sky}}\right]
(C_{l,T} + \Delta_{\rm exp,1})^2 \nonumber \\ \nonumber
{\rm Cov}(C_{l,E},C_{l,E}) & = & \left[\frac{2}{(2l+1)f_{\rm sky}}\right]
(C_{l,E} + \Delta_{\rm exp,2})^2 \\ \nonumber
{\rm Cov}(C_{l,C},C_{l,C}) & = & \left[\frac{2}{(2l+1)f_{\rm sky}}\right]
(C_{l,C}^2 + (C_{l,T}+\Delta_{\rm exp,1})(C_{l,E}+\Delta_{\rm exp,2}))\\
{\rm Cov}(C_{l,T},C_{l,E}) & = & \left[\frac{2}{(2l+1)f_{\rm sky}}\right]
C_{l,C}^2\\ \nonumber
{\rm Cov}(C_{l,T},C_{l,C}) & = & \left[\frac{2}{(2l+1)f_{\rm sky}}\right]
C_{l,C} (C_{l,T}+ \Delta_{\rm exp,1})\\ \nonumber
{\rm Cov}(C_{l,E},C_{l,C}) & = & \left[\frac{2}{(2l+1)f_{\rm sky}}\right]
C_{l,C}(C_{l,E}+ \Delta_{\rm exp,2}),
\end{eqnarray}
where $\Delta_{{\rm exp},i}$ are again experimental experimental errors
related to pixel noise and beamwidth \cite{tth}.

For the purposes of the present paper we assume that
$f_{\rm sky}=1$, $\Delta_{{\rm exp},i}=0$ and $l_{\rm max}=1500$, corresponding
to a full-sky survey up to $l_{\rm max}=1500$, limited only by cosmic variance.
If the temperature power spectrum alone is considered this is not too far
from what can be expected with MAP, whereas PLANCK will likely measure
both temperature and polarization to this accuracy.

As the free cosmological parameters we use the matter density, $\Omega_m$,
the cosmological constant, $\Omega_\Lambda$, the baryon density, $\Omega_b$,
the Hubble parameter, $H_0$, the optical depth to reionization, $\tau$,
and the overall normalization, $Q$. 
Instead of using the spectral index of a featureless power-law spectrum, $n_s$,
as the last free parameter, we bin the spectral index
in $k$-space. In practise we use $N$ bins of equal size in $\log (k)$, from
$k = 1.27 \times 10^{-5} \,\, {\rm Mpc}^{-1}$ to $k = 0.25 \,\, {\rm Mpc}^{-1}$ 
(this covers the entire
range of $k$-space which is visible in the CMBR).
In each bin the power spectrum is assumed to follow a power-law, $P(k)_i \propto
k^{n_s(i)-1}$.
Thus, the parameter vector is
\begin{equation}
\Theta =  \{\Omega_m,\Omega_\Lambda,\Omega_b,H_0,\tau,Q,n_s(1),\ldots,n_s(N)\}
\end{equation}
As the reference model around which to calculate the Fisher matrix, we take the
standard CDM model with, $\Omega=\Omega_m=1$, $\Omega_b=0.05$, $H_0=50 \,\, {\rm km}
\, {\rm s}^{-1} \, {\rm Mpc}^{-1}$, and $\tau=0$. 
The reference model has a power-law initial spectrum
with $n_s(i)=1$ for all $i$.

\begin{figure}[h]
\begin{center}
\epsfysize=10truecm\epsfbox{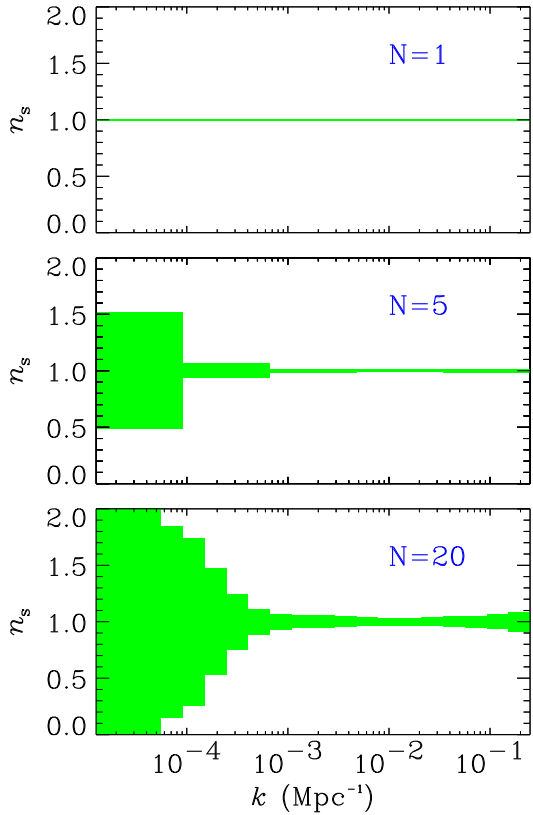}
\end{center}
\vspace*{0.5cm}
\caption{The precision with which the spectral indices in each bin can
be measured, shown for three different numbers of bins ($N=1,5,20$), for the
case where only the temperature anisotropy can be measured.
Shown are the $1\sigma$ error bars around the central value which in all
cases is $n_s=1$.}
\label{fig1}
\end{figure}

In Fig.~1 the precision with which $n_s(i)$ can be measured is shown for different
numbers of bins, for the case where only the temperature power spectrum can
be measured.
Not surprisingly, the precision with which the effective spectral index in
each bin can be measured depends strongly on the number of bins.
For only one bin, $\Delta n_s/n_s = 8.21 \times 10^{-3}$, 
whereas for 20 bins, the smallest $\Delta n_s/n_s$ is $3.49 \times 10^{-2}$.
Thus, there is a trade-off between the resolution in $k$-space and the
resolution in $n_s$. Detecting a low-amplitude narrow feature is not possible,
whereas either a broad low-amplitude or a narrow high-amplitude feature
should be detectable. In agreement with Wang, Spergel and Strauss \cite{wss,wss2}
 we find
that the CMBR data is most sensitive in the range $k \simeq 10^{-3}-10^{-1}
{\rm Mpc}^{-1}$.
If polarization can also be measured, the precision with which the $n_s(i)$
can be measured increases. Fig.~2 shows the same as Fig.~1, but with the
inclusion of polarization. Indeed, the precision on the individual bin
amplitudes increases by a large factor.
For only one bin, $\Delta n_s/n_s = 2.90 \times 10^{-3}$, 
whereas for 20 bins, the smallest $\Delta n_s/n_s$ is now 
$9.49 \times 10^{-3}$.
\begin{figure}[h]
\begin{center}
\epsfysize=10truecm\epsfbox{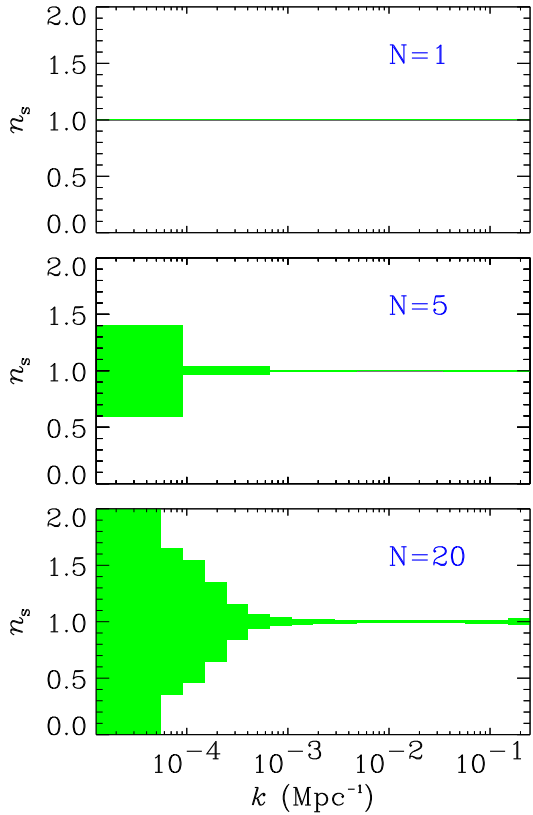}
\end{center}
\vspace*{0.5cm}
\caption{The precision with which the spectral indices in each bin can
be measured, shown for three different numbers of bins ($N=1,5,20$), for the
case where both temperature and polarization can be measured.
Shown are the $1\sigma$ error bars around the central value which in all
cases is $n_s=1$.}
\label{fig2}
\end{figure}

Next, a fundamental question is how much the ability to determine the other
fundamental parameters depends on the number of bins in $k$-space.
Figs.~3 and 4 show the precision in measuring the other cosmological parameters
as a function of $N$, the number of bins.
Fig.~2 is for the case where only the temperature anisotropy can be measured.
For all the cosmological parameters, except $\tau$, there little degeneracy 
between these parameters
and the power spectrum indices, meaning that there is little loss of
ability to pin down these parameters. From $N=10$, the resolution begins to
decrease, although quite slowly, and at $N=30$ it is for instance down by
about 40\% for $\Omega_b$. 
Clearly, the spectral indices, $n_s(i)$, are not very degenerate with
the other cosmological parameters. The reason is that the discontinuous
binning in the power spectrum is very hard to mimic by continuous 
changes in other parameters. This finding is in agreement with Ref.~\cite{wss,wss2}, 
where such a binning was also used. However, there is a marked difference
compared with the findings of Ref.~\cite{soura}. In this work, smoothness of the
power spectrum was enforced by using smooth shape functions peaked at a
given point in $k$-space instead of discontinuous binning. These authors
find that the uncertainty in the other parameters increases very rapidly
with $N$, rising by more than a factor of 10 at $N=30$ \cite{soura}.
However, using smooth functions to describe the power spectrum
could indeed be expected to be able to mimic changes in other parameters
to a much higher degree than our approach. So it is not surprising
that a much higher degree of degeneracy is found.
On the other hand, as will be seen
in the next section, using a discontinuous binning has the very clear
disadvantage that it can be difficult to achieve reasonable likelihood
fits if the number of bins is too small.
\begin{figure}[h]
\begin{center}
\epsfysize=7truecm\epsfbox{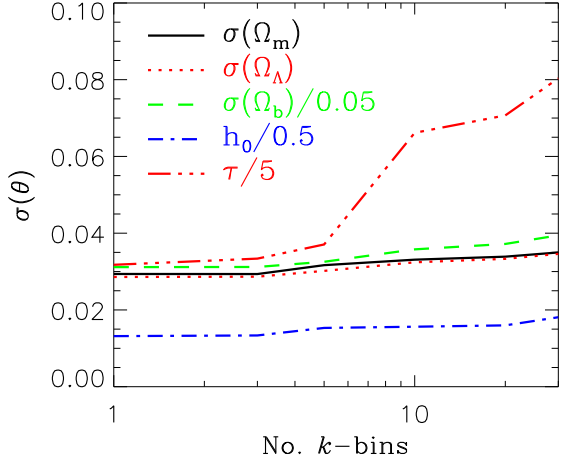}
\end{center}
\caption{The precision with which the other cosmological parameters,
$\Omega_m,\Omega_\Lambda,\Omega_b$ and $H_0$, measured as a
function of the number of bins in $k$-space, for the case where only the temperature 
anisotropy can be measured. The precision is quantified
in terms of the standard error, $\sigma$.}
\label{fig3}
\end{figure}

In line with this discussion, it was shown by Kinney \cite{kinney}
that smooth power spectrum features can mimic variations in the
cosmological parameters almost exactly (to within cosmic variance).
In this work 75 bins in $k$-space were used and the power spectrum
smoothened by cubic spline interpolation, leading to a very large 
degree of degeneracy between the $k$-space power spectrum and
the other cosmological parameters.

The one exception to the above is the optical depth to reionization,
$\tau$. Here there is a very large degree of degeneracy, which increases
rapidly with the number of bins. This effect was also described by
Wang, Spergel and Strauss \cite{wss}. It happens because the effect of
reionization is to suppress the amplitude at small scales by a factor
$e^{-2\tau}$. This can be quite easily mimicked by suppressing the
$k$-space power spectrum at small scales.

However, Fig.~4 shows the case where polarization can also be measured.
In this case, the degeneracy between $\tau$ and the $k$-space spectrum
is broken, and the ability to determine the value of $\tau$ is increased
by a large factor. For the other cosmological parameters, the precision
is also increased, but not as dramatically.

\begin{figure}[h]
\begin{center}
\epsfysize=7truecm\epsfbox{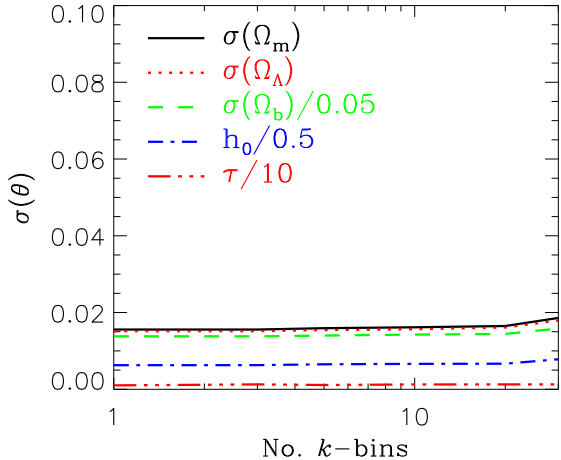}
\end{center}
\caption{The precision with which the other cosmological parameters,
$\Omega_m,\Omega_\Lambda,\Omega_b$ and $H_0$, measured as a
function of the number of bins in $k$-space, for the case where both temperature
and polarization can be measured. The precision is quantified
in terms of the standard error, $\sigma$.}
\label{fig4}
\end{figure}


\section{Numerical Spectrum Reconstruction}

Having done this initial estimate of how precisely the initial
power spectrum can be estimated, it is important to see how such a
power spectrum reconstruction works in practise. 
In order to investigate this, we have produced 10 synthetic spectra based
on a underlying theoretical model. These spectra are calculated assuming
Gaussian errors given by cosmic variance only.
In order to simulate a possible feature in the power spectrum we
introduce a Gaussian ``bump'', so the spectral index has the form
\begin{equation}
n_s = 1 - A \exp\left(-\frac{(\log(k_*)-\log(k_{0,*}))^2}{\alpha}\right),
\end{equation}
where $k_* = k/1 \, {\rm Mpc}^{-1}$.
This bump is characterised by three parameters: $A$, the amplitude,
$k_{0,*}$, the position in $k$-space, and $\alpha$, the width.
For the underlying cosmological model we have chosen the same
standard CDM model as in the last section, so the model parameters
are:
$\Omega=\Omega_m=1$, $\Omega_b=0.05$, $H_0=50 \,\, {\rm km}
\, {\rm s}^{-1} \, {\rm Mpc}^{-1}$, $A=0.7$, $\alpha=0.1$ and
$\log k_{0,*} = -2$.
Thus, the bump has been placed closed to the region where the 
CMBR data is most sensitive. Note that in this section we use only information
related to the temperature power spectrum, not the polarisation
spectrum.

On each of these synthetic spectra we try reconstructing the power spectrum
using the logarithmic binning method introduced in the previous
section. The likelihood optimization algorithm is based on the
simulated annealing principle described in Ref.~\cite{hannestad}.
This method has the advantage that it is very fast for 
highly non-linear optimization
over many-dimensional parameter spaces, exactly the problem that 
is faced here.
In Fig.~3 we show the values extracted by the parameter reconstruction
for three different numbers of bins, $N=5,10$ and 20.

In the $N=5$ case the individual bins are broader than the 
bump to be reconstructed. This shows up in the fact that
no good fit is achieved ($\langle \chi^2_{\rm min} \rangle
/{\rm d.o.f} = 12.7$), and the spectrum reconstruction is quite poor.
The reconstructed spectrum does show some evidence of a lower spectral
index around the position of the bump, but the magnitude is not
correct.
For $N=10$ the reconstruction is already much better and reproduces
the actual slope of the underlying spectrum. However, 
$\langle \chi^2_{\rm min}\rangle/{\rm d.o.f} = 8.06$, so the fit is not very
good in this case either.
In the last case where $N=20$ the reconstructed spectrum again follows
the underlying one reasonably well. However, at $N=20$ the uncertainty on the
individual bin amplitudes is already significantly larger than for
$N=10$. By going to an even higher number of individual bins, the
spectrum bump will be smeared out beyond recognition.
The $\chi^2$-fit
is significantly better than for $N=10$ ($\langle \chi^2_{\rm min}\rangle
/{\rm d.o.f} = 2.84$), although far from what should be expected 
from a ``good'' fit ($\langle \chi^2_{\rm min}\rangle
/{\rm d.o.f} \simeq 1$). 

\begin{figure}[h]
\begin{center}
\epsfysize=10truecm\epsfbox{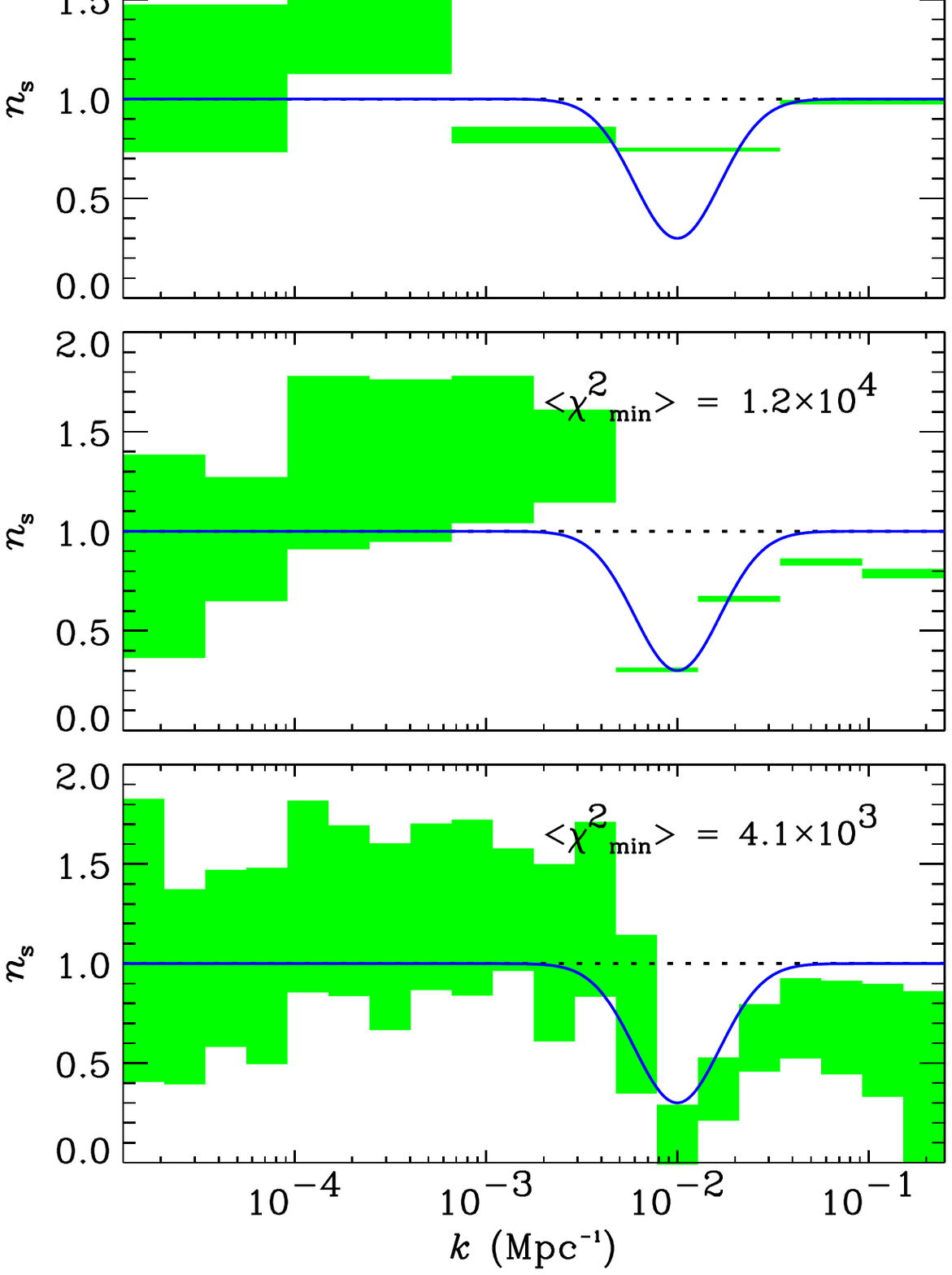}
\end{center}
\vspace*{0.5cm}
\caption{The reconstructed power spectrum from likelihood maximisation
over 10 synthetic spectra. The boxes show the $1\sigma$ error intervals.
The full line shows the underlying $k$-space power spectrum used to
generate the synthetic spectra.}
\label{fig5}
\end{figure}

The reason for these very high $\chi^2$ values
is that it is quite hard to mimic a smooth feature
in the power spectrum with a sequence of discontinuous bins. 
Thus, even though it is indeed possible to map out the shape
of the power spectrum, it is not possible to achieve a good fit
in terms of the likelihood without letting $N \to {\rm d.o.f}$. 
However, going much
beyond $N=20$ increases the uncertainty in the individual bin
amplitudes and doing the likelihood maximization is already
quite demanding numerically at $N=20$.

Finally, with a large number of bins
(10 or 20)
the reconstruction at scales smaller than the bump (large $k$)
is consistently predicting a too low spectral index.
This effect is worrying and to investigate whether it is a generic
feature of any discontinuous binning method, or specific to our 
choice, where each bin has a ``tilt'', $P(k)_i \propto
k^{n_s(i)-1}$, we also do the reconstruction using the same method
as Wang, Spergel and Strauss \cite{wss}. Here, the power in each bin
is assumed to have constant amplitude over the entire bin so that
$P(k)_i = A_i$. We show the reconstruction for the case of $N=20$
in Fig.~6. 
For this case, the reconstruction of the power spectrum bump is
actually better than for the ``tilt'' spectrum reconstruction.
There is no trace of the spuriously low spectral index found
at high $k$. However, the $\chi^2$ is substantially worse for this method,
with $\langle \chi^2_{\rm min}\rangle
/{\rm d.o.f} = 3.45$.
From this, a robust conclusion is that for discontinuous binning 
the obtainable fits are quite poor in terms of $\chi^2$, regardless
of the binning method. However, it is possible to detect features in 
the spectrum by this method.
At least the tilted binning method can show spurious effects, so one
should be careful about making strong conclusions without going
through several different methods of spectrum reconstruction.

\begin{figure}[h]
\begin{center}
\epsfysize=7truecm\epsfbox{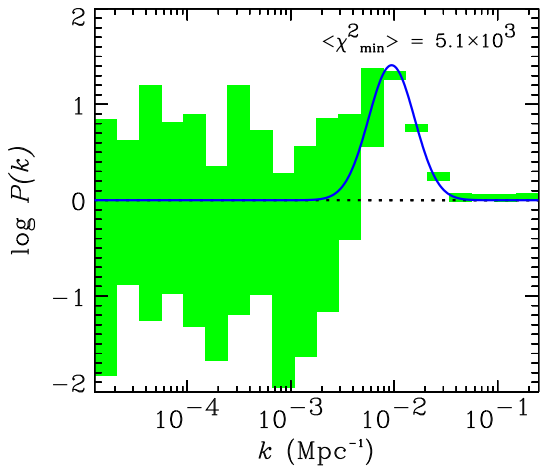}
\end{center}
\vspace*{0.5cm}
\caption{The reconstructed power spectrum from likelihood maximisation
over 10 synthetic spectra, for the case of $N=20$. 
The boxes show the $1\sigma$ error intervals.
The full line shows the underlying $k$-space power spectrum used to
generate the synthetic spectra.}
\label{fig6}
\end{figure}

Just for comparison we have also done a spectrum reconstruction where
the free parameters are those describing a smooth scale invariant
power-law spectrum ($n_s=1$) with a
Gaussian bump: $A, k_{0,*}$ and $\alpha$,
instead of using $n_s(i)$. The values extracted 
from likelihood maximisation on the 10 synthetic spectra are given
in Table I. Clearly, in this case, the true underlying spectrum is
recovered to very good accuracy because the same functional form is
used in the fit as in the underlying power spectrum. 
Also the $\chi^2$ values are completely consistent with what is
expected from a good fit.

So using smooth fitting functions
could have the advantage of being better able to fit smooth
features in the underlying initial power spectrum. However, the
discontinuous binning method has the advantage of being very model
independent. So a possible way to proceed would be to initially
try the discontinuous binning method, and only for refinements go to
more complicated fitting functions.


\section{Discussion}

We have discussed the possibility for measuring the shape of the
primordial power spectrum by using precision measurements of 
CMBR fluctuations. The method employed was to parametrise the spectrum
as a power-law, binned in $k$-space. An analysis of the obtainable
precision in measuring the power spectrum, using a Fisher matrix
analysis, showed that it should be possible to detect features
that are sufficiently broad by using CMBR observations.

However, there are some practical problems involved in this spectrum
reconstruction. If the power spectrum is parametrised by binning
it in $k$-space, it is very difficult to mimic any smooth features
in the underlying power spectrum. Especially if the feature is
comparable in width to the individual bins the reconstructed spectrum
can show spurious features. Nevertheless, by going to a 
sufficiently fine binning
it is possible to map out the general shape of the power spectrum,
although still very difficult to get a good fit in terms of the likelihood.

The fact that it is possible to reconstruct the underlying $k$-space
power spectrum from CMBR observations opens up the possibility of
probing physics at the time of spectrum formation by observing
the universe at CMBR formation ($T \sim 4000$K). 
As was for instance discussed in 
Refs.~\cite{lt94,tw95}, if the primordial fluctuations are produced
during the inflationary epoch, it is possible to reconstruct the
inflationary potential, $V(\phi)$, from the $k$-space spectrum.

However, in addition to that it will be possible to detect 
non-power-law features in the power spectrum using this discrete
binning method. Such features are for instance predicted to occur
in the multiple inflation model of Adams, Ross and Sarkar 
\cite{ars97}. Here, the effect of symmetry breaking during the
inflationary period was discussed. If such spontaneous symmetry
breaking occurs along flat directions, short inflationary periods
will result, leaving distinct non power-law features in the 
power spectrum. Another possible mechanism for producing features
is the resonant production of particles during inflation
discussed by Chung {\it et al.} \cite{chung}.

Note that the CMBR is mainly sensitive to power spectrum at
$k \simeq 10^{-3}-10^{-1} \,\, {\rm Mpc}^{-1}$. Any features outside
this region in $k$-space would not be detectable, so the CMBR can
at most probe a very small region of the inflationary period.
However, as discussed for instance in Ref.~\cite{wss,wss2}, by also
using data from large scale structure (LSS) surveys such as the Sloan 
Digital Sky Survey it is possible to increase the region of 
$k$-space that can be probed. LSS surveys are sensitive
to smaller scales than the CMBR (for instance the Sloan survey 
should probe the region $k \simeq 10^{-2}-1 \,\, {\rm Mpc}^{-1}$),
and by combining MAP/PLANCK with such surveys it should be possible
to have sensitivity in the region $k \simeq 10^{-3}-1 \,\, {\rm Mpc}^{-1}$.
In addition to this, including data from large scale surveys can break
some of the cosmological parameter degeneracies \cite{eisenstein}.
For instance, using
CMBR data alone it is impossible to determine either $\Omega_m$ or
$\Omega_\Lambda$ separately with high precision. The parameter that
can be measured is the total energy density $\Omega=\Omega_m+\Omega_\Lambda$.
However, the LSS surveys are sensitive to a different combination of
$\Omega_m$ and $\Omega_\Lambda$, and by adding this information
the degeneracy between them can be broken \cite{eisenstein}.

\acknowledgements

Use of the CMBFAST code developed by Seljak and Zaldarriaga \cite{SZ96} is 
acknowledged.


\narrowtext
\begin{table}
\caption{The recovered parameters from the fit to a Gaussian bump
overlayed on a scale invariant spectrum. 10 synthetic spectra
were used for the reconstruction.}
\begin{tabular}{lcc}
Parameter & found & expected \cr
\tableline
$\chi^2_{\rm min}$ & $1516.7 \pm 50.4$ & $1496 \pm 54.7$  \\ 
$\alpha$ & $0.1004 \pm 0.0035$ & 0.1 \\
$\log k_{0,*}$ & $-2.0004 \pm 0.0059$ & $-2$ \\
$A$ & $0.6931 \pm 0.003$ & $0.7$
\end{tabular}
\end{table}

\end{document}